\documentclass[aps,pra,twocolumn,superscriptaddress,nofootinbib]{revtex4-1}
\usepackage{amsmath}
\usepackage{amssymb}
\usepackage{graphicx}
\usepackage{mathrsfs}
\usepackage{ntheorem}
\usepackage{mathtools}
\usepackage{enumerate}
\usepackage{enumitem}
\usepackage{times,txfonts}
\usepackage{subfigure}
\usepackage{upgreek}
\usepackage{tikz}
\usepackage{bm}
\usepackage[colorlinks,bookmarksopen,bookmarksnumbered,citecolor=blue, linkcolor=blue, urlcolor=blue]{hyperref}
\newcommand{\ket}[1]{|#1\rangle}
\newcommand{\bra}[1]{\langle #1|}

\newcommand{\abs}[1]{\lvert #1\rvert}

\def\CC{{\rm\kern.24em \vrule width.04em height1.46ex depth-.07ex \kern-.30em C}}
\def\RR{{\rm\kern.24em \vrule width.04em height1.46ex depth-.07ex
\kern-.30em R}}
\def\P{{\rm I\kern-.25em P}}
\begin{document}
\title{Insufficiency of Pure-State Ensembles in Characterizing Transformations of Entangled States under LOCC}

\author{C. L. Liu}
\email{clliu@sdu.edu.cn}
\affiliation{School of Information Science and Engineering, Shandong University, Qingdao 266237, China}
 \author{Baoqing Sun}
  \email{baoqing.sun@sdu.edu.cn}
  \affiliation{School of Information Science and Engineering, Shandong University, Qingdao 266237, China}
\author{D. L. Zhou}
\email{zhoudl72@iphy.ac.cn}
\affiliation{Institute
  of Physics, Beijing National Laboratory for Condensed Matter
  Physics, Chinese Academy of Sciences, Beijing 100190, China}
\affiliation{School of Physical Sciences, University of Chinese
  Academy of Sciences, Beijing 100049, China}
\date{\today}

\begin{abstract}
The conditions for transforming pure entangled states under local operations and classical communication (LOCC) are well understood. A natural question then arises: Can we determine the transformation conditions for mixed entangled states under LOCC based on the properties of their pure-state ensembles? While much effort has been devoted to this issue, in this paper, we rule out this possibility. Our findings address several open questions, including: (i) The conditions \( E_f^{cr}(\rho) \geq E_f^{cr}(\sigma) \) for all convex roof entanglement measures \(E_f^{cr}\) is insufficient to guarantee the existence of an LOCC transformation \(\Lambda^L(\cdot)\) from \(\rho\) to \(\sigma\); and (ii) The inequalities \(\sum_j p_j E(\varphi_j) \geq \sum_l q_l E(\psi_l)\) for all entanglement monotones \(E\) are not sufficient to ensure the existence of an LOCC transformation from \(\{p_j, \ket{\varphi_j}\}\) to \(\{q_l, \ket{\psi_l}\}\).
\end{abstract}
\maketitle
\section{Introduction}
Quantum entanglement, first recognized by Einstein, Podolsky, and Rosen \cite{Einstein}, as well as Schr\"{o}dinger \cite{Schrodinger}, was viewed by Einstein and his collaborators as evidence of the incompleteness of quantum mechanics. However, with Bell's work in 1964 \cite{Bell} and subsequent experiments \cite{Clauser,Zeilinger,Aspect,Clasuer1}, quantum entanglement has been recognized as one of the fundamental features of quantum mechanics. Then, the understanding of quantum entanglement began to shift from a qualitative to a quantitative perspective. It is now recognized as a crucial resource for accomplishing various tasks in quantum information processing \cite{Horodecki,Plenio,Guhne,Dur,Aolita,Chitambar1,Nielsen1}, such as dense coding, teleportation, and reduction of communication complexity. In these applications, a situation of particular interest in quantum information processing consists of two parties that are spatially separated from each other and share a composite system in an entangled state. This setting requires the parties to act only local operations and classical communication (LOCC) on their subsystems \cite{Bennett1}.

A central topic in the theory of quantum entanglement is the transformation of entangled states under LOCC. This investigation was initiated by Lo and Popescu \cite{Lo} and a seminal result in this area was established by Nielsen \cite{Nielsen}, who provided the necessary and sufficient conditions for transforming between two entangled pure states. Specifically, let \(\ket{\psi_1} = \sum_{i=0}^{n-1} \sqrt{\mu_i} \ket{i_A i_B}\) and \(\ket{\psi_2} = \sum_{j=0}^{n-1} \sqrt{\eta_j} \ket{j_A j_B}\) be the Schmidt decompositions of \(\ket{\psi_1}\) and \(\ket{\psi_2}\), respectively. Without loss of generality, we assume \(0 \leq \mu_i \leq \mu_{i+1}\) and \(0 \leq \eta_j \leq \eta_{j+1}\) for all \(i\) and \(j\). Then, \(\ket{\psi_1}\) can be transformed into \(\ket{\psi_2}\) using LOCC if and only if
\begin{eqnarray}
E_l(\psi_1) := \sum_{i=0}^l \mu_i \geq E_l(\psi_2) := \sum_{j=0}^l \eta_j
\end{eqnarray}
holds for all \(0 \leq l \leq n-1\). For simplicity, we denote \(E_l(\psi_1) \geq E_l(\psi_2)\) for all \(l\) by writing \(\uplambda_{\psi_1} \prec \uplambda_{\psi_2}\) \cite{Bhatia,Horn}. Here, $\uplambda_{\psi}$ is the vector of squared Schmidt coefficients of the state $\ket{\psi}$. Subsequently, Jonathan and Plenio \cite{Jonathan} demonstrated that the transformation from \(\ket{\psi}\) to \(\rho\) can be accomplished via LOCC if and only if \(\rho\) has a pure-state ensemble
\(\{p_\alpha, \ket{\varphi_\alpha}\}_{\alpha=1}^N\) such that
\begin{eqnarray}
\sum_{\alpha=1}^N p_\alpha E_l(\ket{\varphi_\alpha}) \leq E_l(\ket{\psi})
\end{eqnarray} hold for all \(l\).

The above results provide a clear understanding of the conditions under which transformations between pure entangled states can be achieved using LOCC. Building on these foundational findings, many researchers have sought to extend the analysis to more general scenarios, including transformations between mixed entangled states as well as between pure-state ensembles \cite{Gour, Jonathan, Li, Ghiu, Li1, Girard, Barnum, Hayden}. On the one hand, Ref. \cite{Jonathan} and related works might lead one to expect that, if a certain inequality is satisfied for all convex-roof entanglement measures, this would be sufficient to guarantee the possibility of an LOCC transformation between two mixed states. Motivated by this suggestion, several studies have explored this direction in depth \cite{Gour, Li, Ghiu, Li1, Barnum, Hayden}, though no definitive conclusion has yet been reached. On the other hand, when considering transformations between pure-state ensembles under LOCC, it has been conjectured \cite{Gour} that a transformation \(T: D_1 \to D_2\) between two probability distributions of bipartite pure states can be implemented via LOCC if and only if \(E(D_1) \geq E(D_2)\) for all entanglement monotones \(E\), where \(D := \{p_j, \ket{\varphi_j}\}\) and \(E(D) := \sum_j p_j E(\ket{\varphi_j})\).

In this paper, we will address these open questions. To tackle these issues, we will first demonstrate that the pure-state ensembles of entangled states are insufficient for characterizing their transformations under LOCC. More precisely, we will demonstrate that there exist two entangled states, \(\rho_{AB}\) and \(\sigma_{AB}\), with \(\{p_j, \ket{\varphi_j}_{AB}\}\) and \(\{q_i, \ket{\psi_i}_{AB}\}\) as arbitrary ensembles of \(\rho_{AB}\) and \(\sigma_{AB}\), respectively, such that it is possible to use an LOCC transformation to convert \(\{p_j, \ket{\varphi_j}_{AB}\}\) into \(\{q_i, \ket{\psi_i}_{AB}\}\), but it is not possible to transform \(\rho_{AB}\) into \(\sigma_{AB}\) using only LOCC (noting that in the conversion from one pure-state ensemble to the other pure-state ensemble, it is assumed that one always knows which pure state one is starting with). Then, we will use this result to study the above two questions. Our results indicate that to provide the necessary and sufficient conditions for the transformation of mixed entangled states under LOCC, we need to introduce additional mathematical tools.

\section{Some Concepts in Entanglement Theory}
To address the above issues clearly, it is helpful to review several related concepts, including separable states, entangled states, separable operations, and LOCC (local operations and classical communication). Separable States: A bipartite quantum state \(\rho_{AB}\) is called separable if it can be expressed as a convex combination of product states
\begin{eqnarray}
\rho_{AB} = \sum_\alpha p_\alpha \rho_\alpha^A \otimes \rho_\alpha^B,
\end{eqnarray}
 where \(\{p_\alpha\}\) is a probability distribution and \(\rho_\alpha^A\) and \(\rho_\alpha^B\) are density operators acting on the Hilbert spaces \(\mathcal{H}_A\) and \(\mathcal{H}_B\), respectively \cite{Werner}. A state is considered entangled if it cannot be represented as a separable state. Separable Operations: A separable operation \(\Lambda^S(\cdot)\) on a bipartite quantum system is a transformation of the form \cite{Rains,Horodecki}
\begin{eqnarray}
\Lambda^S(\rho_{AB}) = \sum_{n=1}^N K_n^A \otimes K_n^B \, \rho_{AB} \, {K_n^A}^\dag \otimes {K_n^B}^\dag,
\end{eqnarray}
where \(\rho_{AB}\) is the initial density operator on the Hilbert space \(\mathcal{H}_A \otimes \mathcal{H}_B\). The Kraus operators \(K_n^A \otimes K_n^B\) are arbitrary product operators that satisfy the trace-preserving condition
\begin{eqnarray}
\sum_n {K_n^A}^\dag K_n^A \otimes {K_n^B}^\dag K_n^B = \mathbb{I}_A \otimes \mathbb{I}_B,
\end{eqnarray}
where \(\mathbb{I}_A\) and \(\mathbb{I}_B\) are the identity operators on parties A and B, respectively. The extension to multipartite systems is straightforward; however, we will focus on the bipartite case here. Therefore, for simplicity, we will omit the indices \(AB\) in the following discussion.

LOCC describes all the transformations that two separated parties (party A and party B) can perform when they apply local operations and have access to classical communication. In general, local operations encompass measurements and operations involving additional local systems \cite{Bennett1}. Classical communication allows the two parties to exchange information over a classical channel. Specifically, consider two subsystems, one located in party A's laboratory and the other in party B's. Suppose party A starts by performing a generalized measurement with outcomes corresponding to Kraus operators \(K_{i_1}\). If the initial state of the system was \(\ket{\psi}\) and party A obtains outcome \(i_1\), the state becomes \(K_{i_1} \otimes \mathbb{I}_B \ket{\psi}\). Party A then communicates the outcome \(i_1\) to party B, who performs a measurement on his subsystem conditioned on \(i_1\), described by Kraus operators \(K_{i_2}^{(i_1)}\). After obtaining outcome \(i_2\), party B informs party A, who then performs another measurement with Kraus operators \(K_{i_3}^{(i_1,i_2)}\). This process can continue for an arbitrary number of rounds. The probabilities of outcomes obtained at each stage must always sum to unity, leading to the conditions$
\sum_{i_n} {K_{i_n}^{(S_n)}}^\dag K_{i_n}^{(S_n)} = \mathbb{I}_A, \quad \text{and} \quad \sum_{i_n} {K_{i_n}^{(S_n)}}^\dag K_{i_n}^{(S_n)} = \mathbb{I}_B,$
where \(S_n\) denotes the collection of indices \(\{i_1, i_2, ..., i_{n-1}\}\) indicating all outcomes obtained in previous measurements.

It has been proved that LOCC is complex and challenging to characterize in its full mathematical detail \cite{Chitambar2}. Furthermore, it is important to note that the set of LOCC operations is a proper subset of separable operations. This means that while all LOCC operations are separable operations, not all separable operations can be achieved through LOCC.

\section{Characterizing Transformations of Entangled States via Pure-state Ensembles}

With these concepts clarified, we are now ready to present the main result of this paper.

\emph{Lemma 1}.--Let \(\rho\) be any \(M \otimes N\) mixed state (without loss of generality, we assume \(M \geq N\)), and let \(\ket{\varphi}\) be any pure entangled state with Schmidt rank \(r\). Then, a necessary condition for achieving the transformation from \(\rho\) to \(\ket{\varphi}\) using LOCC is that \(M \geq 2r\) and \(N \geq r\). (See the Appendix I for the detailed proof.)

We emphasize that if the LOCC in the above lemma is replaced with separable operations, the necessary condition in the above lemma will no longer hold. This can be demonstrated by the following transformation: Consider the initial state
\begin{eqnarray}
\rho= p\ket{\varphi_1}\bra{\varphi_1} + (1-p)\ket{\varphi_2}\bra{\varphi_2}, \label{initialstate}
\end{eqnarray}
where the two states $\ket{\varphi_1}$ and $\ket{\varphi_2}$ above are defined as
$\ket{\varphi_1}=\frac{1}{\sqrt{r}}\left(\ket{01}+\ket{10}+\sum_{j=3}^r\ket{jj}\right)$ and
$\ket{\varphi_2}=\frac{1}{\sqrt{r}}\left(\ket{02}+\ket{20}+\sum_{j=-3}^{-r}\ket{jj}\right)$, respectively.
The final state is given by
\begin{eqnarray}
\ket{\varphi}=\sqrt{\frac{2}{r}}\left(\sqrt{\mu}\nu\ket{00} + \sqrt{\mu}\ket{11} + \frac{1}{\sqrt{2}}\sum_{j=3}^r \ket{jj}\right), \label{finalstate}
\end{eqnarray}
which has a Schmidt rank of \(r\), where \(\mu = \frac{15}{16}\) and \(\nu = \frac{1}{\sqrt{15}}\). Using Lemma 1, it is straightforward to verify that the transformation from \(\rho\) to \(\ket{\varphi}\) via LOCC is impossible, but a separable operation can achieve the transformation from \(\rho\) to \(\ket{\varphi}\). (See the Appendix II for the specific form of the separable operation.)

Lemma 1 and the examples above can be seen as generalizations of the results presented in Ref. \cite{Chitambar}, which compared the differences between using separable operations and LOCC to perform entanglement distillation on 3\(\otimes\)3-dimensional entangled states. Based on these results, we now state the following Theorem 1.

\emph{Theorem 1.}--There exist two entangled states, $\rho$ and $\sigma$, such that, for an arbitrary decomposition $\{p_j,\ket{\varphi_j}\}$ of $\rho$, it is possible to convert each element $\ket{\varphi_j}$ into $\sigma$ using only LOCC, but it is not possible to transform $\rho$ into $\sigma$ using only LOCC.

\emph{Proof}.--First, we consider the case where the initial state is a mixed state \(\rho\) and the final state is a pure state \(\sigma=\ket{\varphi}\bra{\varphi}\). Let \(\{\uplambda_j, \ket{\varphi_j}\}\) be an arbitrary pure-state decomposition of \(\rho\).
We will show that if there exists a separable operation \(\Lambda^{S}(\cdot)\) such that \(\Lambda^S(\rho) = \ket{\varphi}\bra{\varphi}\), then for any state \(\ket{\psi}\) within the span of \(\{\ket{\varphi_j}\}\), i.e., \(\ket{\psi} = \sum_j c_j \ket{\varphi_j}\), we can always construct an LOCC operation \(\Lambda^{L}(\cdot)\) such that
\begin{eqnarray}
\Lambda^{L}(\psi)=\varphi.
\end{eqnarray}
Hereinafter, we will use $\varphi:= \ket{\varphi}\bra{\varphi}$ for any pure state.

To see this, let the separable operation $\Lambda^S(\cdot)$ have the form
$\Lambda^S(\rho) = \sum_n K_n^A \otimes K_n^B \rho {K_n^A}^\dag \otimes {K_n^B}^\dag.$
Then, since pure states are the extreme points of the convex set of states, we have
\begin{equation}
K_n^A \otimes K_n^B \ket{\varphi_j} = \mu_j^n \ket{\varphi}, \label{Kraus_operator}
\end{equation}
for all \(n\) and \(j\). By using Eq. (\ref{Kraus_operator}), we have
\begin{eqnarray}
K_n^A \otimes K_n^B \ket{\psi}= K_n^A \otimes K_n^B \sum_j c_j \ket{\varphi_j} = \sum_j c_j \mu_j^n \ket{\varphi}.
\end{eqnarray}
Since \(\Lambda^S\) is trace preserving, we then obtain
\begin{eqnarray}
\Lambda^S(\psi) = \varphi. \label{sep_trans}
\end{eqnarray}
 By using a result from Ref. \cite{Gheorghiu}, which states that the transformation from $\ket{\psi}$ to $\ket{\varphi}$ can be achieved via some separable operation $\Lambda^S(\cdot)$ if and only if $\uplambda_{\psi} \prec \uplambda_{\varphi}$, we immediately have that there is $\uplambda_{\psi} \prec \uplambda_{\varphi}.$ Furthermore, from Nielsen's theorem \cite{Nielsen}, we obtain that for each $\psi$ there is an LOCC $\Lambda^L(\cdot)$ such that
\begin{eqnarray}
\Lambda^L(\psi)=\varphi. \label{LOCC}
\end{eqnarray}

Next, let us demonstrate that if there exists a separable operation $\Lambda^S(\cdot)$ achieving the transformation from $\rho$ to $\ket{\varphi}$, whereas LOCC cannot, then for any ensemble \(\{p_j, \ket{\varphi_j}\}\) of \(\rho\), even if there are $\uplambda_{\varphi_j}\prec\uplambda_\varphi$, we still cannot guarantee the existence of an LOCC $\Lambda^L(\cdot)$ that transforms \(\rho\) into $\ket{\varphi}$.

Suppose there exists a separable operation $\Lambda^S(\cdot)$ such that $\Lambda^S(\rho) = \ket{\varphi}\bra{\varphi}$, and $\{p_j, \ket{\varphi_j}\}$ is an arbitrary pure-state ensemble of $\rho$, i.e., $\rho= \sum_j p_j \ket{\varphi_j}\bra{\varphi_j}$. According to the arguments around Eq. (\ref{sep_trans}), we immediately obtain that there are $\Lambda^{S}(\psi)=\varphi$ for all states $\ket{\psi}=\sum_jc_j\ket{\varphi_j}$. Furthermore, given any pure-state ensemble $\{q_i,\ket{e_i}\}$ of $\rho$, it can be realized from $\{p_j,\ket{\varphi_j}\}$ by the relation
\begin{eqnarray}
\sqrt{q_i}\ket{e_i}=\sum_jU_{ij}\sqrt{p_j}\ket{\varphi_j},
 \end{eqnarray}
 where $U=[U_{ij}]$ is a unitary matrix \cite{Hughston}. This implies that any state $\ket{\psi}$ within the support of $\rho$ can be transformed into $\ket{\varphi}$ through $\Lambda^S(\cdot)$, where the support of $\rho$ is the vector space spanned by the eigenvectors of $\rho$ with non-zero eigenvalues. Next, according to the arguments around Eq. (\ref{LOCC}), we get that $\ket{\psi}$ can be transformed into $\ket{\varphi}$ using LOCC. This further implies that, although every element in an arbitrary ensemble \( \{p_j, \ket{\varphi_j}\} \) of \( \rho \) can be transformed into \( \ket{\varphi} \) using some LOCC, there exists no LOCC \( \Lambda^L(\cdot) \) that achieves the transformation \( \Lambda^L(\rho) = \varphi \) since \( \rho \) and \( \varphi \) do not satisfy the conditions in Lemma 1.

With the above arguments, let us consider the $\rho_{AB}$ be the state in Eq. (\ref{initialstate}) and the final state $\sigma$ be the state in Eq. (\ref{finalstate}). They satisfy all the properties we discussed above.

Finally, we consider the case where both the initial state and the final state are mixed states. Let the final  state be
\begin{eqnarray}
\sigma=q\ket{\varphi}\bra{\varphi}+(1-q)\ket{\varphi_r}\bra{\varphi_r}, \label{finalstate1}
\end{eqnarray}
where $0\leq q\leq1$ and the state $\ket{\varphi_r}$ is defined as
\begin{eqnarray}
\ket{\varphi_r}=\sqrt{\frac{2}{r}}\left(\sqrt{\mu}\nu\ket{rr} + \sqrt{\mu}\ket{r\text{+}1,r\text{+}1} + \frac{1}{\sqrt{2}}\sum_{j=3}^r \ket{r\text{+}j,r\text{+}j}\right).\nonumber
\end{eqnarray}
Since $\sigma$ and \(\varphi\) can be transformed into each other using LOCC \cite{note1}, it follows directly that \(\rho\) can be transformed into \(\sigma\) through separable operations. However, since $\rho$ cannot be transformed into $\varphi$ by LOCC while $\rho$ and $\varphi$ can be inter-converted by LOCC., it follows that $\rho$ cannot be transformed into $\sigma$ by LOCC. Based on the above arguments, the proof of the theorem is now complete. ~~~~~~~~~~~~~~~~~~~~~~~~~~~~~~~~~~~~~~~~~~~~~~~~~~~~~~~~~~~~~~$\blacksquare$

Theorem 1 tells us that treating a mixed state as a collection of pure states and equating the transformation of mixed states under LOCC to a global transformation between two sets of pure states is incorrect. Thus, to study the conditions for transforming mixed entangled states under LOCC, one must consider properties beyond those of pure-state ensembles. Additionally, Theorem 1 highlights that the operational differences between separable operations and LOCC provide valuable insights for addressing certain problems. In the following, we will use this result to analyze several previously introduced problems.

The first problem is related to the relationship between convex roof entanglement measures and the transformations of mixed entangled states under LOCC, as discussed in Refs. \cite{Ghiu, Li, Jonathan, Li1}. Given that the transformations between pure entangled states and the transformations from pure entangled states to mixed entangled states using LOCC can be characterized via convex roof entanglement measures, some researchers conjecture that the transformations between mixed entangled states under LOCC can also be characterized via convex roof entanglement measures \cite{Ghiu,Li,Jonathan,Li1}. However, as we will demonstrate, this is not the case.

To illustrate this, let us present the concepts of entanglement monotones and convex roof entanglement measures. For a bipartite pure state, assume its Schmidt form is \(\ket{\varphi} = \sum_{i=0}^{n-1} \sqrt{\uplambda_i} \ket{ii}\). For simplicity, we denote \(\bm{\uplambda}_\varphi = (\uplambda_0, \uplambda_1, \ldots, \uplambda_{n-1})^T\) as the Schmidt vector of \(\ket{\varphi}\), where \(0 \leq \uplambda_i \leq \uplambda_{i+1}\). Each entanglement monotone corresponds to a function \(f\) that maps the simplex of probability vectors to real numbers, where \(f\) is both symmetric and concave \cite{Vidal}. Such a function defines an entanglement monotone for pure states \(\ket{\varphi}\), denoted as
\begin{eqnarray}
E_f(\varphi) = f(\bm{\uplambda}_\varphi),
\end{eqnarray}
by evaluating \(f\) on the vector of Schmidt coefficients \(\bm{\uplambda}_\varphi\) of \(\ket{\varphi}\). It has been proven that every entanglement monotone on pure states can be obtained in this manner. Given an entanglement monotone \(E_f\) defined on pure states as \(E_f(\varphi)\), we can extend it to mixed states using the convex roof construction, also known as convex roof entanglement measures \cite{Vidal}
\begin{eqnarray}
E_f^{cr}(\rho) = \inf \sum_i p_i E_f(\varphi_i),
\end{eqnarray}
where the infimum is taken over all ensembles \(\{p_i, \ket{\varphi_i}\}\) such that \(\rho = \sum_i p_i \ket{\varphi_i} \bra{\varphi_i}\). Some well-known entanglement measures, including the entanglement of formation \cite{Bennett1,Wootters}, are defined in this manner. With these notions, we arrive at the following Theorem 2.

\emph{Theorem 2}.--The conditions
\begin{eqnarray}
E_f^{cr}(\rho) \geq E_f^{cr}(\sigma)
\end{eqnarray}
for all convex roof entanglement measures $E_f^{cr}$ are insufficient to guarantee the existence of an LOCC $\Lambda^L(\cdot)$ from $\rho$ to $\sigma$.

\emph{Proof.}--Let \(\rho\) be the state in Eq. (\ref{initialstate}) and \(\sigma\) be the state in Eq. (\ref{finalstate1}). Suppose \(\{p_j, \ket{\varphi_j}\}\) is an arbitrary pure-state ensemble that generates \(\rho\). Based on the arguments in the Theorem 1, it is straightforward to verify that we can achieve the transformations from \(\ket{\varphi_j}\) to \(\ket{\varphi}\) in Eq. (\ref{finalstate}) for all \(j\) using LOCC. Furthermore, using the result that any ensemble decomposition of \(\rho = \{q_i, \ket{e_i}\}\) can be obtained from \(\{p_j, \ket{\varphi_j}\}\) through the relation \(\sqrt{q_i}\ket{e_i}= \sum_j U_{ij} \sqrt{p_j} \ket{\varphi_j}\), where \(U = [U_{ij}]\) represents a unitary matrix, and applying the arguments in Theorem 1, it follows that the transformation from \(\ket{e_i}\) to \(\varphi\) can also be accomplished for all \(i\) using LOCC. Based on these results, we can deduce that for any convex roof entanglement measure \(E_f^{cr}\), it follows that \(E_f^{cr}(\ket{\varphi_j}) \geq E_f^{cr}(\ket{\varphi})\) for all \(j\). Since we have $E_f^{cr}(\ket{\varphi})=E_f^{cr}(\sigma)$ for all $E_f^{cr}$, it follows that there are
\begin{eqnarray}
\sum_iq_iE_f^{cr}(\ket{e_i})\geq E_f^{cr}(\sigma)
\end{eqnarray}
for any pure-state ensemble $\{q_i,\ket{e_i}\}$ of $\rho$. This further implies that
\begin{eqnarray}
E_f^{cr}(\rho) \geq E_f^{cr}(\sigma)
\end{eqnarray}
for all convex roof entanglement measures $E_f^{cr}$. Nevertheless, according to Theorem 1, it is impossible to achieve the transformation from \(\rho\) to \(\sigma\) using LOCC. Therefore, the condition \(E_f^{cr}(\rho) \geq E_f^{cr}(\sigma)\) for all convex roof entanglement measures \(E_f^{cr}\) are not sufficient to guarantee the existence of an LOCC \(\Lambda^L(\cdot)\) that transforms \(\rho\) to \(\sigma\). This concludes the proof of the theorem. ~~~~~~~~~~~~~~~~~~~~~~~~~~~~~~~~~~~~~~~~~~~~~~~~~~~~~~~~~~~~~~~~~~~~~~~~~~~~~~~~~~~~$\blacksquare$

Based on Theorem 2 above, we obtain that we cannot determine the conditions for the transformation between mixed states under LOCC solely by using convex roof entanglement measures. It is noteworthy that this result is closely related to another research topic. This topic primarily investigates whether there exists a set of entanglement measures $\{E_\mu\}$ that constitutes a complete set \cite{Gour, Datta}. Here, completeness means that if all entanglement measures in $\{E_\mu\}$ satisfy the inequality $E_\mu(\rho) \geq E_\mu(\sigma)$, then there exists an LOCC $\Lambda^L(\cdot)$ that realizes the transformation from $\rho$ to $\sigma$. Theorem 2 indicates that even if $\{E_\mu\}$ includes all convex roof entanglement measures, it still does not satisfy completeness.

The second open question concerns the conditions for the transformation between two pure-state ensembles using LOCC. The transformation from one pure-state ensemble \( D_1 := \{ p_j, \ket{\varphi_j} \} \) to another \( D_2 := \{ q_l, \ket{\psi_l} \} \) is a probabilistic process in which Alice and Bob initially share \( D_1 \), and use LOCC to convert it into \( D_2 \). Specifically, for each ensemble element \( \ket{\varphi_j} \) in \( D_1 \), Alice and Bob apply a transformation \( T_j \) that outputs the state \( \ket{\psi_l} \) with conditional probability \( q_{l|j} \), such that the overall probability of obtaining \( \ket{\psi_l} \) in \( D_2 \) is \( q_l = \sum_j p_j q_{l|j} \). This process is a valid LOCC transformation if a set of LOCC steps can achieve the conversion with the required probabilities. In Ref. \cite{Gour}, it is conjectured that: A transformation \( T: D_1 \to D_2 \) between two probability distributions \cite{note} of bipartite mixed states can be realized by LOCC iff \( E(D_1) \geq E(D_2) \) for all entanglement monotones \( E \). Here, for an ensemble \( D := \{ p_j, \ket{\varphi_j} \} \), \( E(D) \) is defined as \( E(D) := \sum_j p_j E(\ket{\varphi_j}) \).

Using the results above, we demonstrate that this conjecture is invalid, which leads us to the following Theorem 3:

\emph{Theorem 3.}--The conditions
\begin{eqnarray}
\sum_j p_j E(\varphi_j) \geq \sum_l q_l E(\psi_l)
\end{eqnarray}
for all entanglement monotones $E$ are not sufficient to guarantee the existence of a transformation of a pure-state ensemble $\{p_j, \ket{\varphi_j}\}$ to another $\{q_l, \ket{\psi_l}\}$ can be achieved by LOCC.

\emph{Proof.}--We will use proof by contradiction to demonstrate the above theorem. Assume that if \(\sum_j p_j E(\varphi_j) \geq \sum_l q_l E(\psi_l)\) for all entanglement monotones \(E\), then the transformation of \(\{p_j, \ket{\varphi_j}\}\) to \(\{q_l, \ket{\psi_l}\}\) can be achieved by some LOCC \(\Lambda^L(\cdot)\). According to the Lemma 1 above, let us choose the initial pure-state ensemble as $\{p,\ket{\varphi_1};1\text{-}p,\ket{\varphi_2}\}$ and the target pure-state ensemble as $\{q, \ket{\psi_1};1\text{-}q, \ket{\psi_2}\}$, where the initial states are
\begin{eqnarray}
\ket{\varphi_1}=\frac{1}{\sqrt{r}}(\ket{01}+\ket{10}+\sum_{j=3}^r\ket{jj})\nonumber
\end{eqnarray}
 and
\begin{eqnarray}
\ket{\varphi_2}=\frac{1}{\sqrt{r}}(\ket{02}+\ket{20}+\sum_{j=-3}^{-r}\ket{jj}),\nonumber
\end{eqnarray}
and the final states are
\begin{eqnarray}
\ket{\psi_1}= \sqrt{\frac{2}{r}}(\sqrt{\mu}\nu\ket{00} + \sqrt{\mu}\ket{11} + \frac{1}{\sqrt{2}}\sum_{j=3}^r \ket{jj}),\nonumber
\end{eqnarray}
and
\begin{eqnarray}
\ket{\psi_2}=\sqrt{\frac{2}{r}}(\sqrt{\mu}\nu\ket{rr} + \sqrt{\mu}\ket{r\text{+}1,r\text{+}1} + \frac{1}{\sqrt{2}}\sum_{j=3}^r \ket{r\text{+}j,r\text{+}j}).\nonumber
\end{eqnarray}
According to the Nielsen's theorem \cite{Nielsen}, we get that the transformations from $\ket{\varphi_1}$ to $\ket{\psi_1}$ and $\ket{\psi_2}$ to $\ket{\psi_r}$ can both be achieved using LOCC. This indicates that for all entanglement monotones, there are $E(\ket{\varphi_j}) \geq E(\ket{\psi_1})$ and $E(\ket{\varphi_j}) \geq E(\ket{\psi_2})$ for $j=1,2$.

On the other hand, if there is an LOCC, say $\Lambda^L(\cdot)$, achieving the transformation from the ensemble $\{p,\ket{\varphi_1};1\text{-}p,\ket{\varphi_2}\}$ to $\{q,\ket{\psi_1};1\text{-}q,\ket{\psi_2}\}$, then according to the Theorem 1, there are $\Lambda^L(\ket{\varphi_j})=\ket{\psi_1}$ and $\Lambda^L(\ket{\varphi_j})=\ket{\psi_2}$ for all $j=1,2$. This further implies that \(\sum_j p_j E(\varphi_j) \geq \sum_l q_l E(\psi_l)\) for all entanglement monotones \(E\). According to the assumption, we can thus transform $\rho= p\ket{\varphi_1}\bra{\varphi_1}+(1\text{-}p)\ket{\varphi_2}\bra{\varphi_2}$ into $\sigma=q\ket{\psi_1}\bra{\psi_1}+(1\text{-}q)\ket{\psi_2}\bra{\psi_2}$ using LOCC.  However, this contradicts the result presented in Theorem 1, completing the proof of the theorem. ~~~~~~~~~~~~~~~~~~~~~~~~~~~~~~~~~~~~~~~~~~~~~~~~~~~~~~~~~~~~~~~~~~~~~~~~~~~~~~$\blacksquare$

\section{Conclusions}
In summary, we have addressed the question: Can the transformation conditions of mixed entangled states under LOCC be determined based on the properties of their pure-state ensembles? Theorem 1 rules out this possibility, demonstrating that treating a mixed state as a collection of pure states and equating the transformation of mixed states under LOCC to a global transformation between two sets of pure states is incorrect. Building on this result, we resolve two conjectures, presented in Theorems 2 and 3. While we show that both conjectures are incorrect, these results are still significant in a positive way, as they reveal that the direction taken by previous approaches was not fruitful. To study the conditions for transforming mixed entangled states under LOCC, one must consider properties beyond those of pure-state ensembles.

\begin{acknowledgments}
This work is supported by  National Key Research and Development Program of China (Grants No. 2021YFA1402104 and No. 2021YFA0718302) and National Natural Science Foundation of China (Grant No. 12075310 and 12405019)
\end{acknowledgments}

\section*{Appendixes}

\subsection*{Appendix I. The proof of Lemma 1}

\emph{Lemma 1}.--Let $\rho$ be any $M\otimes N$ (without loss of generality, we assume $M\geq N$) mixed state and let $\ket{\varphi}$ be any pure entangled state with its Schmidt rank being $r$. Then it is possible to deterministically transform $\rho$ into $\ket{\varphi}$ only if $M\geq2r$ and $N\geq r$.

\emph{Proof.}--First, we show that for any state \(\ket{\phi}\) in the span of \(\{\ket{\varphi_j}\}\), where \(\{\ket{\varphi_j}\}\) are the eigenvectors of \(\rho\), the Schmidt rank of \(\ket{\phi}\) is at least \(r\).

Let \(\rho = \sum_j p_j \ket{\varphi_j} \bra{\varphi_j}\) be the spectral decomposition of \(\rho\), and suppose there exists an LOCC \(\Lambda\) such that \(\Lambda(\rho) = \sum_n K_n^A \otimes K_n^B \rho {K_n^A}^\dag \otimes {K_n^B}^\dag\), which achieves the transformation \(\Lambda(\rho) = \varphi\). Since pure states are the extremal points of the set of density matrices, it follows directly that \(\Lambda(\varphi_j) = \varphi\) for all \(j\). This implies that
\begin{eqnarray}
K_n^A \otimes K_n^B \ket{\varphi_j} = c_j^n \ket{\varphi}, \label{transformation}
\end{eqnarray}
for some constant \(c_j^n\). From these conditions, we can conclude that for any state \(\ket{\phi}\) in the span of \(\{\ket{\varphi_j}\}\), i.e., \(\ket{\phi} = \sum_j \mu_j \ket{\varphi_j}\), we have \(\Lambda(\phi) = \varphi\). To see this, starting from Eq. (\ref{transformation}), we derive
\begin{eqnarray}
\Lambda(\phi) &&=\sum_n K_n^A \otimes K_n^B \ket{\phi} \bra{\phi} {K_n^A}^\dag \otimes {K_n^B}^\dag\nonumber\\
&&=\sum_n \sum_{j,k} \mu_j \mu_k^* c_j^n {c_k^n}^* \ket{\varphi} \bra{\varphi}. \label{transformation-1}
\end{eqnarray}
On one hand, by using the trace-preserving property of \(\Lambda\), i.e., \(\sum_n {K_n^A}^\dag K_n^A \otimes {K_n^B}^\dag K_n^B = \mathbb{I}_A \otimes \mathbb{I}_B\) and Eq. (\ref{transformation}), we obtain
\begin{eqnarray}
\sum_n \bra{\varphi_k} {K_n^A}^\dag K_n^A \otimes {K_n^B}^\dag K_n^B \ket{\varphi_j} = \sum_n {c_k^n}^* c_j^n. \label{orthogonal-1}
\end{eqnarray}
On the other hand, since \(\{\ket{\varphi_j}\}\) are the eigenvectors of \(\rho\), we have
\begin{eqnarray}
\sum_n \bra{\varphi_k} {K_n^A}^\dag K_n^A \otimes {K_n^B}^\dag K_n^B \ket{\varphi_j} = \bra{\varphi_k} \mathbb{I}_A \otimes \mathbb{I}_B \ket{\varphi_j} = \delta_{kj}. \label{orthogonal-2}
\end{eqnarray}
From Eqs. (\ref{orthogonal-1}) and (\ref{orthogonal-2}), we obtain
\begin{eqnarray}
\sum_n {c_k^n}^* c_j^n = \delta_{kj}. \label{condition-1}
\end{eqnarray}
Additionally, since \(\ket{\phi} = \sum_j \mu_j \ket{\varphi_j}\) is a state, we also have the normalization condition
\begin{eqnarray}
\sum_j \abs{\mu_j}^2 = 1. \label{condition-2}
\end{eqnarray}
Substituting Eqs. (\ref{condition-1}) and (\ref{condition-2}) into Eq. (\ref{transformation-1}), we obtain
\begin{eqnarray}
\Lambda(\phi) = \varphi.
\end{eqnarray}
This further implies, by applying Nielsen's theorem from Ref. \cite{Nielsen}, that the Schmidt rank of \(\ket{\phi}\) is at least \(r\).

Second, we show that it is possible to deterministically transform $\rho$ into $\ket{\varphi}$ only if $M\geq2r$ and $N\geq r$.

The condition \(N \geq r\) is obvious. Thus, we only need to prove that $M\geq2r$. According to the definition of LOCC, it is understood that $\Lambda$ is implemented alternately by party A and party B with the help of classical communication. Assume that the operation has reached a certain stage, and the corresponding Kraus operator at this stage is \(K_{i_n}^{(S_n)}\otimes K_{i_n}^{(S_n)}\).  Next, we will use induction to prove that \( K_{i_n}^{(S_n)} \otimes K_{i_n}^{(S_n)} \ket{\varphi_j} \neq 0 \) if and only if \( K_{i_n}^{(S_n)} \otimes K_{i_n}^{(S_n)} \ket{\varphi_k} \neq 0 \) for all \( k \neq j \). Suppose that the next operation is to be performed by party A. Then, for any possible Kraus operators performed by party A, \( K_{j} \), the corresponding accumulated operation will become $\left(A_j\otimes \mathbb{I}\right)\left(K_{i_n}^{(S_n)} \otimes K_{i_n}^{(S_n)}\right)$. According to the inductive hypothesis, if \( K_{i_n}^{(S_n)} \otimes K_{i_n}^{(S_n)} \ket{\varphi_j} \neq 0 \), then \( K_{i_n}^{(S_n)} \otimes K_{i_n}^{(S_n)} \ket{\varphi_k} \neq 0 \) for all \( k \neq j \). Since \(\Lambda\) is a deterministic transformation, it follows that all \( K_{i_n}^{(S_n)} \otimes K_{i_n}^{(S_n)} \ket{\varphi_j} \) have Schmidt rank at least \( r \). Therefore, we further conclude that if $\left(A_j\otimes \mathbb{I}\right)\left(K_{i_n}^{(S_n)} \otimes K_{i_n}^{(S_n)}\right)\ket{\varphi_j}\neq0$ for all $j$.

Next, consider any branch in the protocol that transforms \( \ket{\varphi_j} \) to \(\ket{\varphi}\). As just proven, \( \ket{\varphi_k} \) must also be transformed to \( \ket{\varphi} \) along this branch. Since \( \{ \ket{\varphi_j} \} \) are orthogonal but ultimately become the same state, there must exist some round \( N \) on this branch such that \( \{ \ket{\varphi^{N-1}_j} \} \) are not parallel, but \( A\otimes\mathbb{I}\ket{\varphi^{N-1}_j} = c_k(A\otimes\mathbb{I}) \ket{\varphi^{N-1}_k}\), where \( c_k\neq0\) are some complex scalar. Here, \( \{\ket{\varphi_j^{N-1}}\} \) denote the resultant state of the system originally in state \( \{ \ket{\varphi_j} \} \) after round \( N-1 \), and it is assumed, without loss of generality, that Alice makes the \( N \)-th round measurement. Because all \( \{\ket{\varphi_j^{N-1}}\} \) can be transformed into \( \ket{\varphi}\) by the same LOCC protocol, it follows by the linearity of the protocol that any superposition of \( \{\ket{\varphi_j^{N-1}}\} \) must be entangled. However, it is impossible that \( A\otimes\mathbb{I}\left(\ket{\varphi^{N-1}_j}-c_k \ket{\varphi^{N-1}_k}\right)= 0 \), since \( A \) is of rank at least $r$, which leads to a contradiction. This completes the proof of the lemma.  ~~~~~~~~~~~~~~~~~~~~~~~~~~~~~~~~~~~~~~~~~~~~~~~~~~~~~~~~~~~~~~~~~~~~~~~~~~~~~~$\blacksquare$

\subsection*{Appendix II. The Counterexample in the Main Text}

We will show that if the LOCC in the lemma 1 above is replaced with separable operations, the necessary conditions in the lemma will no longer hold. This can be demonstrated with the following counterexample.

Let us consider the following initial state with its dimension being $(2r-1)\otimes(2r-1)$
\begin{eqnarray}
\rho=p\ket{\varphi_1}\bra{\varphi_1}+(1-p)\ket{\varphi_2}\bra{\varphi_2}
\end{eqnarray}
where
$\ket{\varphi_1}=\frac1{\sqrt{r}}\left(\ket{01}+\ket{10}+\sum_{j=3}^r\ket{jj}\right)~\text{and}~
\ket{\varphi_2}=\frac1{\sqrt{r}}\left(\ket{02}+\ket{20}+\sum_{j=-3}^{-r}\ket{jj}\right)$
and the final state has a Schmidt rank of $r$
\begin{eqnarray}
\ket{\varphi}=\sqrt{\frac2r}\left(\sqrt{\mu}\nu\ket{00}+\sqrt{\mu}\ket{11}+\frac1{\sqrt{2}}\sum_{j=3}^r\ket{jj}\right)
\end{eqnarray}
It is direct to verify that the transformation from $\rho$ to $\ket{\varphi}$ can be achieved using the following separable operation
\begin{eqnarray}
\Lambda(\cdot)=\sum_{j=1}^{17}K_j\cdot K_j^\dag
\end{eqnarray}
where the Kraus operators $K_j$ with $j=1,...,17$ are given as
\begin{widetext}
\begin{eqnarray}
K_1&&=\left(\sqrt{\mu\nu}\ket{0}\bra{0}+\sqrt{\mu}\ket{1}\bra{1}+\frac1{\sqrt[4]{2}}\sum_{j=3}^r\ket{j}\bra{j}\right)
\otimes\left(\sqrt{\nu}\ket{0}\bra{1}+\ket{1}\bra{0}+\frac1{\sqrt[4]{2}}\sum_{j=3}^r\ket{j}\bra{j}\right)\nonumber\\
K_2&&=\left(\sqrt{\mu}\ket{1}\bra{0}+\sqrt{\mu\nu}\ket{0}\bra{1}+\frac1{\sqrt[4]{2}}\sum_{j=3}^r\ket{j}\bra{j}\right)
\otimes\left(\ket{1}\bra{1}+\sqrt{\nu}\ket{0}\bra{0}+\frac1{\sqrt[4]{2}}\sum_{j=3}^r\ket{j}\bra{j}\right)\nonumber\\
K_3&&=\left(\sqrt{\mu\nu}\ket{0}\bra{0}+\sqrt{\mu}\ket{1}\bra{2}+\frac1{\sqrt[4]{2}}\sum_{j=-3}^{-r}\ket{-j}\bra{j}\right)
\otimes\left(\sqrt{\nu}\ket{0}\bra{2}+\ket{1}\bra{0}+\frac1{\sqrt[4]{2}}\sum_{j=3}^r\ket{-j}\bra{j}\right)\nonumber\\
K_4&&=\left(\sqrt{\mu}\ket{1}\bra{0}+\sqrt{\mu\nu}\ket{0}\bra{2}+\frac1{\sqrt[4]{2}}\sum_{j=3}^r\ket{-j}\bra{j}\right)
\otimes\left(\ket{1}\bra{2}+\sqrt{\nu}\ket{0}\bra{0}+\frac1{\sqrt[4]{2}}\sum_{j=3}^r\ket{-j}\bra{j}\right)\nonumber\\
K_5&&=\ket{1}\bra{1}\otimes\left(\sqrt{1-2\mu\nu}\ket{1}\bra{1}+\ket{2}\bra{2}\right),~~~
K_6=\ket{2}\bra{2}\otimes\left(\sqrt{1-2\mu\nu}\ket{2}\bra{2}+\ket{1}\bra{1}\right)\nonumber\\
K_7&&=\sqrt{1-4\mu\nu}\ket{0}\bra{0}\otimes\ket{0}\bra{0},\nonumber\\
K_8&&=\sum_{j=3}^r\ket{j}\bra{j}\otimes\sum_{j=-3}^{-r}\ket{j}\bra{j},~~~
K_9=\sum_{j=-3}^{-r}\ket{j}\bra{j}\otimes\sum_{j=3}^{r}\ket{j}\bra{j}\nonumber\\
K_{10}&&=\sum_{j=3}^r\ket{j}\bra{j}\otimes\ket{2}\bra{2},~~~
K_{11}=\ket{2}\bra{2}\otimes\sum_{j=3}^r\ket{j}\bra{j}\nonumber\\
K_{12}&&=\sum_{j=-3}^{-r}\ket{j}\bra{j}\otimes\ket{1}\bra{1},~~~
K_{13}=\ket{1}\bra{1}\otimes\sum_{j=-3}^{-r}\ket{j}\bra{j}\nonumber\\
K_{14}&&=\left(\sqrt{1-\frac{\mu\nu+\mu}{\sqrt{2}}}(\ket{0}\bra{0}+\ket{1}\bra{1})\right)\otimes\sum_{j=3}^{r}\ket{j}\bra{j},~~~
K_{15}=\sum_{j=3}^{r}\ket{j}\bra{j}\otimes\left(\sqrt{1-\frac{\nu+1}{\sqrt{2}}}(\ket{0}\bra{0}+\ket{1}\bra{1})\right)\nonumber\\
K_{16}&&=\left(\sqrt{1-\frac{\mu\nu+\mu}{\sqrt{2}}}(\ket{0}\bra{0}+\ket{2}\bra{2})\right)\otimes\sum_{j=-3}^{-r}\ket{j}\bra{j},~~~
K_{17}=\sum_{j=-3}^{-r}\ket{j}\bra{j}\otimes\left(\sqrt{1-\frac{\nu+1}{\sqrt{2}}}(\ket{0}\bra{0}+\ket{2}\bra{2})\right)\nonumber
\end{eqnarray}
\end{widetext}
with $\mu=\frac{15}{16}$ and $\nu=\frac1{\sqrt{15}}$.


\begin{thebibliography}{99}
\bibitem{note1} The result that there exists an LOCC protocol achieving the transformation from \(\varphi\) to \(\sigma\) is based on the findings in Ref. \cite{Jonathan}. On the other hand, the transformation from \(\sigma\) to \(\varphi\) can be achieved using the operation \(\Lambda(\cdot) = pI \otimes I \cdot I \otimes I + (1 - p)E \otimes E \cdot E \otimes E\), where \(E = \sum_{j} \ket{j}\bra{j+r}\), which can also be realized via LOCC.
\end{thebibliography}
\end{document}